\documentclass[11pt]{article}
\usepackage{amsfonts}
\usepackage{graphics,amssymb,amsthm,amsmath,epsf}
\usepackage{graphicx}
\usepackage{subfigure}

\numberwithin{equation}{section}

\allowdisplaybreaks

\begin{document}
\title{Comparison of some Different Methods for Hypothesis Test of Means of
  Log-normal Populations}
\author{Saba Aghadoust, Kamel Abdollahnezhad\thanks{Corresponding: k.abdollahnezhad@gu.ac.ir  } , Farhad Yaghmaei, Ali Akbar Jafari  \ \\
{\small Department of Statistics, Golestan University, Gorgan, Iran}\\
{\small Department of Statistics, Yazd University, Yazd,  Iran}\\}
\date{}
\maketitle

\begin{abstract}
The log-normal distribution is used to describe the positive data, that it has skewed distribution with small mean and large
variance. This distribution has application in many sciences for example medicine, economics, biology
and alimentary science, ect. Comparison of means of several log-normal populations
 always has been in focus of researchers, but the test statistic are not easy to derive or extremely
 complicated for this comparisons. In this paper, the different methods exist for this testing that we can point out F-test,
likelihood ratio test, generalized p-value approach and computational approach test. In this line
with help of simulation studies, in this methods  we compare and evaluate
 size and power test.
\end{abstract}

\section{Introduction}
The skewed distributions are particularly common when mean values are small, variances are large and values cannot be negative (for example lengths of latent periods of infectious diseases), and often closely fit the log-normal distribution. The log-normal distribution has been widely used in medical, biological and economic studies, where data are positive and have a right-skewed distribution

Let${\ X}_{ij}$ be random sample from $k$ independent log-normal distributions, i.e.

\[Y_{ij}={\log  \left(X_{ij}\right)\ }\sim N\left({\mu }_i,{\sigma }^2_i\right),\ \ i=1,.\ .\ .\ ,k;;\ j=1,.\ .\ .\ ,n_i.\]

Also, let ${\varphi }_i=E\left(X_{ij}\right)={\rm exp}({\mu }_i+\frac{1}{2}{\sigma }^2_i)$ denote the mean of the $i$-th population. Suppose we are interested in testing

\begin{equation}
H_0:\ {\varphi }_1={\varphi }_2=.\ .\ .\ ={\varphi }_k{\rm \ \ \ \ \ \ \ vs}{\rm .\ \ \ \ \ \ \ \ \ }H_A:{\rm Not\ all}{\mathbf \ }{\rm the}{\rm \ }{\varphi }_is{\rm \ }{\rm are}{\rm \ }{\rm equal}.
\end{equation}
Then the testing problem (1.1) is equivalent to testing
\begin{equation}
H_0: \eta_i=\eta,{\rm \ \ \ \ \ \ \ \ \ vs}.\ \ \ \ \ \ \ \ \ H_A{\rm :\ Not\ all\ }{\rm the}{\rm \ }{\eta }_is{\rm \ }{\rm are}{\rm \ }{\rm equal}
\end{equation}
where $\eta_i={\log  \left({\varphi }_i\right)\ }={\mu }_i+\frac{1}{2}{\sigma }^2_i$, and $\eta$ is unspecified.

It is well-known that the maximum likelihood estimators (MLE's) for ${\mu }_i$  maximum likelihood estimation and ${\sigma }^2_i$ are ${\overline{Y}}_i$ and $S^2_i$,  respectively, where
\begin{equation}
{\overline{Y}}_i=\frac{1}{n_i}\sum^{n_i}_{j=1}{Y_{ij}}{\rm \ \ \ \ \ \ \ \ and\ \ \ \ \ \ \ \ \ }S^2_i=\frac{1}{n_i}\sum^{n_i}_{j=1}{{\left(Y_{ij}-{\overline{Y}}_i\right)}^2},
\end{equation}
Therefore, the MLE of $\eta_i$ is  ${\widehat{\eta }}_i={\widehat{\mu }}_i+\frac{1}{2}{\widehat{\sigma }}^2_i$ that it has approximately normal distribution with mean ${\eta }_i$ and variance $v_i=\frac{{\sigma }^2_i}{n_i}+\frac{\left(n_i-1\right){\sigma }^4_i}{2n^2_i}$.

\section{The CAT}

Pal et al. (2007) introduced the CAT in a general setup. Suppose$\ X_1,X_2,\dots ,X_n$ is a random sample from a population with the known density function $f\left(x| \lambda \right)$ with ${\mathbf \lambda }=\left(\theta,\psi \right)$, where $\theta$ is the parameter of interest and $\psi $ is the nuisance parameter. To test $H_0~:\theta=\theta_0$ against a suitable alternative $H_1\ $at level $a$, the general methodology of the CAT for testing is given through the following steps:

\begin{enumerate}
\item  First obtain $\widehat{{\mathbf \lambda }}{\mathbf =}\left(\hat{\theta},\widehat{\psi }\right)$, the MLE of ${\mathbf \lambda }$.

\item  Assume that $H_0$ is true, i.e., set $\theta=\theta_0$. Then find the MLE of $\psi $ from the data. Call this as the `restricted MLE' of $\psi $ under $H_0$ and denote by ${\widehat{\psi }}_{RML}$.

\item   Generate artificial sample $Y_1,Y_2,\dots ,Y_n$ from $f(x|\theta_0,{\widehat{\psi }}_{RML})$ a large number of times (say, $M$ times). For each of these replicated samples, recalculate the MLE of ${\mathbf \lambda }=\left(\theta,\psi \right)$. Retain only the component that is relevant for $è$.Let these recalculated MLE values of $è$be ${\hat{\theta}}_1,{\hat{\theta}}_2,\dots ,{\hat{\theta}}_M$.

\item  Let ${\hat{\theta}}_{(1)}<{\hat{\theta}}_{(2)}<\dots <{\hat{\theta}}_{(M)}$ be the ordered values of ${\hat{\theta}}_l,\ 1\le l\le M$.

\item  (i) For testing $H_0~:\theta=\theta_0$ against $H_1~:\theta<\theta_0$, define ${\widehat{\theta }}_L={\widehat{\theta }}_{(\alpha M)}$. Reject $H_0$ if $\hat{\theta}<{\widehat{\theta }}_L$ and accept $H_0$ otherwise. Alternatively, calculate the p-value as
\[p=\frac{1}{M}\sum^M_{l=1}{I({\hat{\theta}}_l<\hat{\theta})}.\]
\end{enumerate}

(ii) For testing $H_0~:\theta=\theta_0$ against $H_1~:\theta>\theta_0$, define ${\widehat{\theta }}_U={\widehat{\theta }}_{((1-\alpha )M)}$. Reject $H_0$ if $\hat{\theta}>{\widehat{\theta }}_U$ and accept $H_0$ otherwise. Alternatively, calculate the p-value as
\[p=\frac{1}{M}\sum^M_{l=1}{I({\hat{\theta}}_l>\hat{\theta})}.\]

(iii) For testing $H_0~:\theta=\theta_0$ against $H_1~:\theta<\theta_0$, define ${\widehat{\theta }}_L={\widehat{\theta }}_{((\alpha /2)M)}$ and ${\widehat{\theta }}_U={\widehat{\theta }}_{((1-\alpha /2)M)}$. Reject $H_0$ if $\hat{\theta}<{\widehat{\theta }}_L$ or $\hat{\theta}>{\widehat{\theta }}_U$ and accept $H_0$ otherwise. Alternatively, calculate the p-value as:
\[p=2{\min  \left(p_1,1-p_1\right)\ },\]
where $p_1=\frac{1}{M}\sum^M_{l=1}{I({\hat{\theta}}_l<\hat{\theta})}.$

Now implement the CAT for testing the equality of several log-normal means. To apply our proposed CAT, we first need to express $H_0$  in term of a suitable scalar $\theta$. Define
\[\theta=h\left({\mu }_i;;{\sigma }^2_i\right)=\sum^k_{i=1}{\frac{1}{v_i}{\left({\eta }_i-\overline{\eta }\right)}^2}=\sum^k_{i=1}{\frac{{\eta }^2_i}{v_i}}-\frac{{\left(\sum^k_{i=1}{\frac{{\eta }_i}{v_i}}\right)}^2}{\sum^k_{i=1}{\frac{1}{v_i}}},\]
where $\overline{\eta }={\left(\sum^k_{i=1}{\frac{1}{v_i}}\right)}^{-1}\sum^k_{i=1}{\frac{{\eta }_i}{v_i}}$. It is seen that the hypothesis in (1.2) is equivalent to
\[H^*_0:\theta=0\ \ \ \ \ \ \ \ \ vs\ \ \ H^*_A:\theta>0.\]

If we apply the steps of CAT, then we have the following steps for testing the equality of means of several log-normal distributions:

1) Obtain ${\widehat{\mu }}_i={\overline{X}}_i$and ${\widehat{\sigma }}^2_i=S^2_i$, $i=1,\dots ,k$, and calculate $\widehat{\theta }=h\left({\widehat{\mu }}_i;;{\widehat{\sigma }}^2_i\right)$.

2) Assume that $H_0$ in (1.2) is true, i.e., set ${\mu }_i=\eta -\frac{1}{2}{\sigma }^2_i,\ 1=i=k$. The likelihood function $($under $H_0)\ $is a function of $\left(\eta ,{\sigma }^2_1{,\dots ,\sigma }^2_k\right)$ only. The restricted MLE's of $\eta ,{\sigma }^2_1{,\dots ,\sigma }^2_k$, denoted by ${\widehat{\eta }}_{RML},{\widehat{\sigma }}^2_{i\left(RML\right)},\ 1=i=k$, are found using numerical methods, see [6]. Define ${\widehat{\mu }}_{i\left(RML\right)}={\widehat{\eta }}_{RML}-\frac{1}{2}{\widehat{\sigma }}^2_{i\left(RML\right)}$

3) Generate artificial sample $Y_{i1},.\ .\ .\ ,Y_{in_i}\left(={{\mathbf Y}}_{{\mathbf i}},say\right)$ independent random sample from $N\left({\widehat{\mu }}_{i\left(RML\right)},{\widehat{\sigma }}^2_{i\left(RML\right)}\right)$. Repeat this process $M$ times. In the $l$-th replication ($1=l=M)$ based on $Y^{\left(l\right)}_i$ get the MLE's of  ${\mu }_i$ and ${\sigma }^2_i$ by (1.3) and call them as ${\widehat{\mu }}^{(l)}_{0i}$ and ${\widehat{\sigma }}^{2(l)}_{0i}$. Then recalculate $\widehat{\theta }$ as ${\widehat{\theta }}_{0l}=h({\widehat{\mu }}^{\left(l\right)}_{0i},{\widehat{\sigma }}^{2(l)}_{0i})$.

4) Order the ${\widehat{\theta }}_{0l}$ values as ${\widehat{\theta }}_{0(1)}={\widehat{\theta }}_{0\left(2\right)}=.\ .\ .\ ={\widehat{\theta }}_{0\left(M\right)}$.

5) Let ${\widehat{\theta }}_U={\widehat{\theta }}_{0(\left(1-\alpha \right)M)}$ and reject $H_0$ if $\hat{\theta}>{\widehat{\theta }}_U$ and accept $H_0$ otherwise. Alternatively, if the value $p=\frac{1}{M}\sum^M_{l=1}{I\left({\widehat{\theta }}_{0l}>{\widehat{\theta }}_{ML}\right)}$ is smaller than the nominal level $a$, then reject $H_0$.

\end{document}